\documentclass[epj]{svjour}
\usepackage{amsmath,amssymb}
\usepackage{graphicx}

\begin{document}

\title{Monte-Carlo Simulations of Electron Channeling: a Bent (110) Channel in Silicon}
\author{Andriy Kostyuk}                     % Do not remove
\institute{65933 Frankfurt am Main, Germany}

\abstract{
Results obtained with a new Monte-Carlo code ChaS for channeling of 855 MeV electrons 
along the crystallographic plane (110) in a bent silicon crystal are presented. 
The dependence of the dechanneling length and the asymptotic acceptance of the channel
on the crystal bending is studied.
}

\maketitle
%

%\begin{keyword}

%\PACS 61.85.+p \sep 02.70.Uu \sep 41.75.Fr


%\end{keyword}

\section{Introduction}

In this article, channeling of ultrarelativistic electrons in a bent 
planar channel is studied using the Monte-Carlo method.

Channeling has been intensively investigated since the sixties years of the last century.
This phenomenon 
is observed if particles enter a single crystal at a small
angle with respect to a major crystallographic direction \cite{Lindhard}.
Due to the electrostatic potential of the crystal constituents, 
charged projectiles move preferably along the 
corresponding crystallographic planes or axes following their shape.

This suggested the idea \cite{Tsyganov1976} to use crystals with bent 
crystallographic planes to steer high-energy charged particle beams.
Since its successful experimental verification \cite{Elishev1979},
this idea has been inducing growing interest
to practical applications of 
channeling and related phenomena 
(e.g.  volume capture \cite{Andreev1982capture} and volume reflection \cite{Taratin1987reflection}).

In particular, bent crystals are replacing huge dipole magnets
 in the Institute for High Energy Physics in Prot\-vino (Russia)
where they are used for beam extraction and deflection \cite{Afonin2005}. 
A series of experiments on the bent crystal deflection and collimation of proton and heavy ion beams were
performed in other laboratories 
\cite{Arduini1997,Carrigan1999,Fliller2006,Strokov2007,Scandale2008,Scandale:2011za} (see also a
recent review \cite{Scandale:2012zz} and references therein).
It was proposed to extract particles from the beam halo at Large Hadron Collider using bent crystals \cite{Uggerhoj2005}.
The possibility of deflecting positron \cite{Bellucci2006} and electron \cite{Strokov2007,Strokov2006,Scandale2009zz}
beams has been studied as well.

Another interesting application of the channeling phenomenon is 
the crystalline undulator, a novel 
source of hard electromagnetic radiation.
A single crystal with periodically bent crystallographic planes can force 
channeling particles to move along  nearly sinusoidal trajectories
and radiate in the hard x-ray and gamma-ray frequency range 
\cite{Kaplin1980,Baryshevsky1980}. The extremely strong
electrostatic fields inside the crystal are able
to steer the particles much more effectively than even the most advanced
superconductive magnets. Due to this fact, the period of a crystalline undulator
can be made very short (up to several microns\footnote{Recently, it was found 
\cite{Kostyuk:2012mk} that even smaller crystalline undulator
periods (hundreds of nanometers) are feasible and advantageous.}) 
and, therefore, the energy of emitted photons can be in the
range of several MeV or even higher. 

Using positron beams
with the crystalline undulator is preferable \cite{Korol:1999im},
because positrons have much larger dechanneling length than 
electrons.  
On the other hand, electron beams are more easily available and are usually
of higher quality and intensity. For this reason, 
the electron based crystalline undulator \cite{Tabrizi:2006yi,Tabrizi:2007tv}
may have some practical advantages and, therefore, it deserves a thorough 
analysis.

Understanding the behavior of charged particles in a bent crystal channel
is a necessary step towards developing a comprehensive theory of the crystalline 
undulator. In the present paper, channeling of electrons in a bent silicon
crystal is studied  using a new Monte-Carlo 
code ChaS (\textbf{Cha}nneling \textbf{S}imulator).
The code enables one to simulate channeling of charged particles
and analyze their trajectories. 
In contrast to
most of other channeling codes 
\cite{Artru:1990nz,Biryukov:1995hv,Bogdanov_Mathematica_2010,Guidi_2010,Korol:2001ir,Saitoh_code_1985,Taratin_code_1979},
the underlying algorithm is not 
based on the continuous potential approximation. Instead, binary
collisions of the projectile with crystal constituents, nuclei and electrons,
are simulated. This feature may be especially beneficial in the case of negatively 
charged projectiles, which channel in the vicinity of the atomic nuclei, where using 
the continuous potential approximation is not well justified.

The binary collision algorithm is not new in the channeling physics. In fact, it has 
been actively used in the field since 1960s 
\cite{Robinson1963,Kudrin_code_1973,Andersen:1979qs,Bak:1984fw,Smulders_code_1987}.
But the previously existing codes consider binary collisions of
the projectile with the crystal atoms as whole ignoring incoherent collisions with atomic
electrons. In contrast, the algorithm of ChaS takes into account collisions of the projectile 
with target electrons as well as with nuclei. This is important for positrons as well as for
electrons. The absolute contribution of the incoherent scattering by target electrons is expected
to be larger in the case of negatively charged projectiles, that have to cross the crystal plane in the
process of channeling where the electron density is higher than average. But its relative
contribution is larger for positively charged projectiles, that channels between the planes and,
therefore, are being incoherently scattered mostly by target electrons.

The scope of the present paper is restricted to the analysis of channels 
with constant curvature. 
The results on
channeling of 855 MeV electrons in a straight and bent single crystals of silicon along 
the plane (110) are presented. This plane is chosen for the analysis because it can be 
deformed by growing of Si$_{1-x}$Ge$_x$ crystals \cite{Breese97}
with a varying Ge content $x$ \cite{MikkelsenUggerhoj2000,Kostyuk:2013kk}.
The electron energy corresponds to the conditions of the channeling experiments
at Mainz Microtron (Germany) \cite{Backe2008,Backe2011}. 
The results will help one to estimate the reasonable parameters 
of periodically bent crystal channels and, therefore,
will facilitate future simulations and experimental study 
of the electron-based crystalline undulator.
Additionally, the results can be useful to developers of crystalline devices
for extraction, bending and collimation of electron beams.

\section{The Physical Model and the Simulation Algorithm}
\label{Model}

A detailed description of the
physical model and the algorithm that are implemented in the  
Monte Carlo code ChaS can be found in \cite{straight}.
In the present section,
the basic ideas of the model are briefly reviewed.  
The part 
of the algorithm that is relevant to the simulation of
a bent crystal is considered in greater details.

The model is optimized for studying the interaction of ultrarelativistic projectiles
with single crystals. Due to the high speed of the projectile, its interaction 
time with a crystal atom is  very short.  
The motion of the atomic electrons as well as the thermal motion of the atomic nuclei
during the interaction time can be neglected. Therefore, the crystal is represented 
as  a set of static charges. The atomic nuclei are `frozen' at random positions in 
the vicinity of nodes of the crystal lattice. The probability distribution of the position
of the nucleus relative to the node is approximated by a three dimensional Gaussian 
distribution.  The variance of the distribution is equal to the squared amplitude of 
thermal vibrations of the crystal atoms $a(T)$.
The same value as in \cite{straight}, $a(T) = 0.075$~{\AA},
is used in the present calculations.
It corresponds to the room temperature \cite{Crystallography}.

The probability distribution of electrons in the crystal is approximated by, a 
spherically symmetric distribution of $Z$ electrons around each atom 
($Z$ is the atomic number). 
The radial dependence of the distribution is chosen in such a way that it
reproduces  on average the electrostatic potential of the atom in 
Moli\`ere's approximation \cite{Moliere}. 

The code performs 3D simulation of the projectile motion in the crystal.
The coordinate frame is defined in such a way that the  axis $z$ coincides with the
beam direction. The tangent plane to the bent crystallographic plane (110)
at the crystal entrance is parallel to the coordinate 
plane $(xz)$ and, consequently, it is orthogonal to the axis $y$. 

The orientation of the plane is carefully chosen to avoid collinearity of major crystal
axes with the coordinate axis $z$ (otherwise axial channeling would be simulated instead of the planar 
one). The axes are considered to be 'major' if the maximum distance
between the corresponding atomic strings in the plane (110) exceeds the  amplitude of 
thermal vibrations $a(T)$ by the factor of $3$ at least. The orientation of the plane is done in such a way 
that the angle between such axes and 
the coordinate axis $z$ is not smaller than $\sim 10$ mrad.

The interaction of the projectile with a crystal constituent is considered as a 
classical scattering in a Coulomb field of a static point-like charge.
Electrons and nuclei are taken into account if they belong to a lattice node located within 
a cylinder of the radius $40 a_\mathrm{TF}$ around the projectile trajectory 
($a_\mathrm{TF}$ is the Thomas-Fermi radius of the atom). 
Initially, the straight line along the direction of the projectile momentum at the point 
of entering the crystal is taken as the axis of the cylinder.
The length of the cylinder $\Delta z_\mathrm{c}$
is approximately 200 \AA. 
When the projectile approaches the end of the cylinder, a new cylinder is built as 
an extension of the previous one. The axis of the new cylinder is parallel to the  
new particle momentum. The procedure is repeated until the projectile reaches the end of the 
crystal.
As a result, the cylinders form a `pipe' filled with the crystal lattice and the 
particle moves in the middle of this `pipe'.
 
In the case of a bent crystal with channel shape defined by the function $y_\mathrm{B}(z)$, 
the procedure of \cite{straight} is modified in the following way. 
If $(x_\mathrm{p},y_\mathrm{p},z_\mathrm{p})$ are the coordinates of the projectile, and  
$(x_\mathrm{c},y_\mathrm{c},z_\mathrm{c})$ is the endpoint of the cylinder axis: 
\begin{eqnarray}
z_\mathrm{c} &=& z_\mathrm{p}+\Delta z_\mathrm{c}  \nonumber \\
x_\mathrm{c} &=& x_\mathrm{p} + \Delta z_\mathrm{c} p_{x}/p_{z} \\ 
y_\mathrm{c} &=& y_\mathrm{p} + \Delta z_\mathrm{c} p_{y}/p_{z}. \nonumber
\end{eqnarray}
First, an undistorted lattice is built around the straight line segment connecting the points 
$(x_\mathrm{p},y_\mathrm{p}-y_\mathrm{B}(z_\mathrm{p}),z_\mathrm{p})$ and 
$(x_\mathrm{c},y_\mathrm{c}-y_\mathrm{B}(z_\mathrm{c}),z_\mathrm{c})$. Then each lattice 
node is shifted according to the bending profile: 
\begin{equation}
(x_\mathrm{n},y_\mathrm{n},z_\mathrm{n}) 
\rightarrow (x_\mathrm{n},y_\mathrm{n}+y_\mathrm{B}(z_\mathrm{n}),z_\mathrm{n})
\end{equation}
As a result, 
the trajectory becomes surrounded by a  `pipe' filled by the bent crystal lattice.

It is convenient to characterize the curvature of the bent crystal by the centrifugal 
parameter $C$ \cite{Korol:2004ug}.
Let $U(y)$ be the potential energy of the projectile in the field of 
straight crystal averaged over the coordinates $x$ and $z$ at fixed $y$.
Then the centrifugal parameter is defined as\footnote{
The continuous potential is not used in the simulation algorithm of the code
ChaS. It is utilized only for the definition of the centrifugal parameter $C$.}
\begin{equation}
C = \frac{F_\mathrm{c.f.}}{U^{\prime}_{\max}}.
\label{C}
\end{equation}
Here $F_\mathrm{c.f.}$ is the centrifugal force acting on the projectile in the bent channel and 
$U^{\prime}_{\max}$ is the maximum derivative of the average potential energy with respect to $y$,
i.e. the maximum force acting on the projectile in the average potential. 
The centrifugal force is related 
to the projectile energy $E$ and to the bending radius $R_{C}$ via $F_\mathrm{c.f.}=E/R_{C}$. Therefore,
expression (\ref{C}) can be rewritten as 
\begin{equation}
C = \frac{E}{R_{C} U^{\prime}_{\max}}.
\label{CRC}
\end{equation}

The shape of a channel with a constant bending radius
is defined by the equation
\begin{equation}
y_\mathrm{B}(z) = R_{C} - \sqrt{R_{C}^2-z^2}.
\end{equation}
In this case, the potential energy has to be averaged over the surface 
\begin{equation}
y=y_\mathrm{B}(z)+\upsilon
\label{upsilon}
\end{equation}
with fixed $\upsilon$. 

With account for centrifugal effects, 
the potential energy of the particle has the form
\begin{equation}
U_\mathrm{c.f.}(\upsilon) = U(\upsilon) - \upsilon F_\mathrm{c.f.} .
\label{Ucf}
\end{equation}
Because the crystal deformation is not very large,
the modification of the electrostatic part
of the potential energy can be neglected. 
Therefore, it is assumed that the function $U(\upsilon)$ in (\ref{Ucf}) does not depend
on the crystal curvature. This is the reason, why $U^{\prime}_{\max}$ in (\ref{C}) refers
to the straight channel.

\begin{figure}[htp]
\resizebox{0.95\columnwidth}{!}{%
\includegraphics{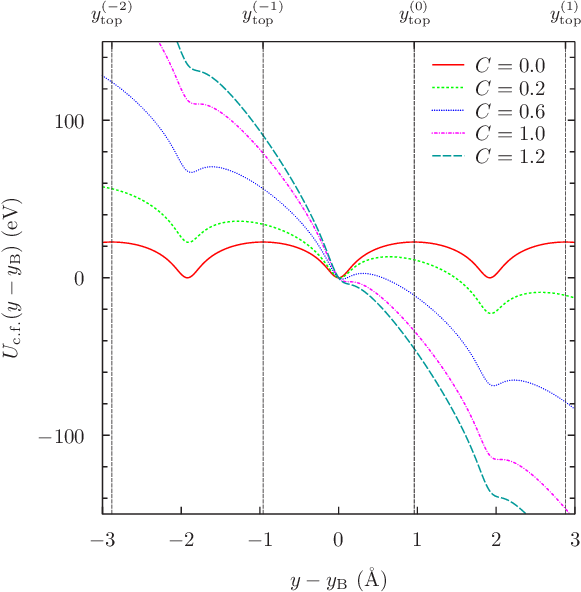}
}
\caption{The average potential energy of the projectile in the field of bent silicon crystal planes (110)
with account for the centrifugal effects. The centrifugal parameter $C$ is defined by equation (\ref{C}).}
\label{PlotUcf}
\end{figure}

The average potential energy of the projectile in the field of bent crystallographic planes (110)
with account for the centrifugal effects is shown in figure \ref{PlotUcf}.
It is seen from the figure and can be deduced from 
(\ref{C}) and (\ref{Ucf})  that 
the value $C=0$ corresponds to the straight crystal and $C=1$ is the critical value at which 
the potential barriers between the crystal channels disappear.

The channel boundaries for a straight crystal are defined as  the planes satisfying the equation
$y = y_\mathrm{top}^{(n)}$,
where $y_\mathrm{top}^{(n)}$ is the coordinate of the potential energy maximum between the $n$-th
and $(n+1)$-st channel. 
In the case of nonzero $C$, these planes become bent surfaces following the shape of the 
channel: 
\begin{equation}
y = y_\mathrm{top}^{(n)}+y_\mathrm{B}(z).
\label{chbound}
\end{equation}
In this case, the channel boundaries 
do not coincide with the maxima of the potential energy that are shifted due to the centrifugal effects 
(see figure \ref{PlotUcf}).

\section{Simulations}

The simulations were performed 
for E=855 MeV electrons channeling along the plane (110) for a number of different
values of $C$ in the range from $C=0$ to $C=1.5$ (see table \ref{tab}). 

The case of an ideal zero-emittance beam entering the crystal 
strictly parallel to the coordinate axis $z$ was simulated, i.e.
the projectiles had zero transverse momentum at the entrance of the crystal.
The initial transverse position of the projectile was chosen randomly, homogeneously distributed
within a rectangular region. The $y$ dimension of the rectangle was exactly equal to the channel 
width and the $x$ dimension  was of the same order of magnitude.
Then the trajectory of the particle was simulated. 
The simulation of the trajectory was
terminated if the particle went through the crystal: $z > L_\mathrm{c}$, or if the 
deviation of the projectile from its initial direction became too large: 
$| \vec{p}_{\perp} | / p_{z} > 0.06 \gg \theta_\mathrm{L}$. Here $\vec{p}_{\perp}$ and $p_{z}$ 
are, respectively, the transverse and the longitudinal momenta and $\theta_\mathrm{L}$ is the 
critical (Lindhard's) angle, $\theta_\mathrm{L} \approx 0.26 \cdot 10^{-3}$ rad
for the channel (110) in silicon at $E=855$ MeV.

Then the simulated trajectories were analyzed and the segments corresponding to channeling 
and dechanneling modes were determined.
At the entrance point of the crystal, the projectile was assumed to be in the channeling mode.
Regardless of the value of its transverse energy,
the particle was considered to be channeling until it crossed the channel boundary
(\ref{chbound}). 

The crystal dimensions along the beam direction $L_\mathrm{c}$
was chosen sufficiently large, 
so that practically all trajectories crossed the channel boundary
at least ones, i.e all particles were already dechanneled at 
$z=L_\mathrm{c}$. The rechanneling process (see \cite{straight} for a detailed
discussion) takes place also in a bent channel. But it is irrelevant 
for the present analysis. The numerical values of $L_\mathrm{c}$ for 
each $C$ as well as the numbers 
of simulated trajectories $N_{0}$ are listed in table \ref{tab}.

\begin{table}
\caption{The parameters used in the simulations:
the centrifugal parameter $C$, 
the crystal length $L_\mathrm{c}$,
and the number of simulated trajectories $N_{0}$.
The obtained values of the dechanneling length $L_\mathrm{d}$ and
the asymptotic acceptance $A_\mathrm{d}$ (see section \ref{definitionLd})
are listed in the last two columns together with their statistical errors.
\label{tab}
}
\begin{tabular}{|c|r|r|r|r|}
\hline
  $C$    &  $L_\mathrm{c}$  ($\mu$m) 
                     & \multicolumn{1}{c|}{$N_{0}$}          
                                             & \multicolumn{1}{c|}{$L_\mathrm{d}$ ($\mu$m)} 
                                                                   &  \multicolumn{1}{c|}{$A_\mathrm{d}$} \\
\hline
 0.000   & 150 \ \ \ &  $ 2.0 \cdot 10^{5} $ & $ 8.401 \pm 0.047 $ & $  0.97 \pm 0.01 $  \\
 0.025   & 130 \ \ \ &  $ 2.5 \cdot 10^{4} $ & $ 8.127 \pm 0.099 $ & $  0.92 \pm 0.02 $  \\
 0.050   & 120 \ \ \ &  $ 2.5 \cdot 10^{4} $ & $ 7.417 \pm 0.091 $ & $  0.88 \pm 0.02 $  \\
 0.075   & 110 \ \ \ &  $ 2.5 \cdot 10^{4} $ & $ 6.664 \pm 0.010 $ & $  0.83 \pm 0.02 $  \\
 0.100   & 100 \ \ \ &  $ 2.5 \cdot 10^{4} $ & $ 5.995 \pm 0.063 $ & $  0.79 \pm 0.01 $  \\
 0.150   & 90  \ \ \ &  $ 2.5 \cdot 10^{4} $ & $ 5.183 \pm 0.066 $ & $  0.68 \pm 0.01 $  \\
 0.200   & 70  \ \ \ &  $ 2.5 \cdot 10^{4} $ & $ 4.298 \pm 0.066 $ & $  0.63 \pm 0.02 $  \\
 0.250   & 60  \ \ \ &  $ 2.5 \cdot 10^{4} $ & $ 3.729 \pm 0.056 $ & $  0.57 \pm 0.01 $  \\
 0.300   & 50  \ \ \ &  $ 2.5 \cdot 10^{4} $ & $ 3.304 \pm 0.070 $ & $  0.51 \pm 0.02 $  \\
 0.400   & 35  \ \ \ &  $ 2.5 \cdot 10^{4} $ & $ 2.594 \pm 0.052 $ & $  0.45 \pm 0.02 $  \\
 0.500   & 30  \ \ \ &  $ 5.0 \cdot 10^{4} $ & $ 1.988 \pm 0.038 $ & $  0.43 \pm 0.02 $  \\
 0.600   & 25  \ \ \ &  $ 5.0 \cdot 10^{4} $ & $ 1.500 \pm 0.018 $ & $  0.43 \pm 0.01 $  \\
 0.700   & 20  \ \ \ &  $ 5.0 \cdot 10^{4} $ & $ 1.176 \pm 0.018 $ & $  0.43 \pm 0.01 $  \\
 0.800   & 15  \ \ \ &  $ 1.0 \cdot 10^{5} $ & $ 0.894 \pm 0.011 $ & $  0.44 \pm 0.01 $  \\
 0.900   & 12  \ \ \ &  $ 1.0 \cdot 10^{5} $ & $ 0.640 \pm 0.008 $ & $  0.48 \pm 0.01 $  \\
 0.950   & 12  \ \ \ &  $ 1.0 \cdot 10^{5} $ & $ 0.502 \pm 0.007 $ & $  0.57 \pm 0.02 $  \\
 0.975   & 12  \ \ \ &  $ 1.0 \cdot 10^{5} $ & $ 0.440 \pm 0.006 $ & $  0.67 \pm 0.02 $  \\
 1.000   & 10  \ \ \ &  $ 2.0 \cdot 10^{5} $ & $ 0.378 \pm 0.005 $ & $  0.84 \pm 0.04 $  \\
 1.010   & 10  \ \ \ &  $ 2.0 \cdot 10^{5} $ & $ 0.349 \pm 0.003 $ & $  0.99 \pm 0.03 $  \\
 1.025   & 8   \ \ \ &  $ 2.0 \cdot 10^{5} $ & $ 0.319 \pm 0.003 $ & $  1.14 \pm 0.04 $  \\
 1.050   & 6   \ \ \ &  $ 2.0 \cdot 10^{5} $ & $ 0.264 \pm 0.002 $ & $  1.85 \pm 0.05 $  \\
 1.100   & 5   \ \ \ &  $ 2.0 \cdot 10^{5} $ & $ 0.195 \pm 0.002 $ & $  4.13 \pm 0.16 $  \\
 1.150   & 4   \ \ \ &  $ 2.0 \cdot 10^{5} $ & $ 0.149 \pm 0.002 $ & $ 10.48 \pm 0.61 $  \\
 1.200   & 3   \ \ \ &  $ 2.0 \cdot 10^{5} $ & $ 0.120 \pm 0.001 $ & $ 26.40 \pm 1.75 $  \\
 1.250   & 3   \ \ \ &  $ 2.0 \cdot 10^{5} $ &       --            &      --             \\
 1.300   & 3   \ \ \ &  $ 2.0 \cdot 10^{5} $ &       --            &      --             \\
 1.500   & 3   \ \ \ &  $ 2.0 \cdot 10^{5} $ &       --            &      --             \\
\hline 
\end{tabular}
\end{table}

\section{The Dechanneling Length and the Asymptotic Acceptance of the Channel}
\label{definitionLd}

To make a quantitative assessment of the particle dechanneling process, one needs a definition 
of the dechanneling length that would be suitable for the Monte Carlo approach. In this section,
the definition of the dechanneling length proposed in \cite{straight} is reviewed and another
useful quantity, the {\it asymptotic acceptance}, is introduced.

Let $z_\mathrm{d1}$ be the point of the first crossing of
the channel boundary by the projectile trajectory.
The quantity $N_\mathrm{ch0}(z)$ is defined as the number of trajectories
that satisfy the condition $z_\mathrm{d1} > z$. In other words,
this is the number of projectiles (among the total number $N_{0}$)
that passed the distance from their entrance into the crystal to the point $z$ in the 
channeling regime and dechannel 
at some further point. The length $L(z)$ is the average distance from the point $z$ to the 
first dechanneling point:
\begin{equation}
L(z) = \frac{\sum_{k=1}^{N_\mathrm{ch0}(z)} (z_\mathrm{d1}^{(k)} - z) }{N_\mathrm{ch0}(z)} .
\label{Lz}
\end{equation}
The sum in the numerator is taken over trajectories satisfying the condition $z_\mathrm{d1} > z$.

\sloppy
Generally speaking, $L(z)$ depends not only on $z$, but also on the initial angular 
distribution of the particles. However, it has been demonstrated in \cite{straight} 
for strait channels 
that $L(z)$ reaches an asymptotic value at sufficiently large $z$.
This value depends neither on $z$ nor on the initial angular distribution. 
It will shown in the next section that $L(z)$ preserves this property 
in bent channels as well.
The dechanneling length $L_\mathrm{d}$ is defined in the Monte Carlo procedure
as the value of 
$L(z)$ given by (\ref{Lz}) in the region where it ceases to depend on $z$.
The number of channeling particles $N_\mathrm{ch0}(z)$ decreases exponentially
in this region 
\begin{equation}
N_\mathrm{ch0}(z) \asymp N_{0} A_\mathrm{d} \exp \left(- z / L_\mathrm{d} \right).
\label{Nch0_diff_asymp}
\end{equation}
Here $N_{0}$ is the total number of simulated trajectories.

Due to the exponential asymptote (\ref{Nch0_diff_asymp}), the quantity
\begin{equation}
A(z) = \frac{N_\mathrm{ch0}(z)}{N_{0}}  \exp \left(z / L_\mathrm{d} \right) 
\label{Az}
\end{equation}
should also approach a constant value $A_\mathrm{d}$ at sufficiently large $z$. The quantity $A_\mathrm{d}$ will 
be called
the {\it asymptotic acceptance} of the channel. Similar quantity can be defined also within the diffusion approach 
\cite{BiryukovChesnokovKotovBook}.
In contrast to the dechanneling length $L_\mathrm{d}$, the asymptotic 
acceptance depends on the initial transverse momentum distribution of the particles in the beam.

The values of $L_\mathrm{d}$ and $A_\mathrm{d}$ calculated for the simulated trajectories are 
listed in table \ref{tab}. 

\section{Analysis of the Results}

\begin{figure}[htp]
\resizebox{0.95\columnwidth}{!}{%
  \includegraphics{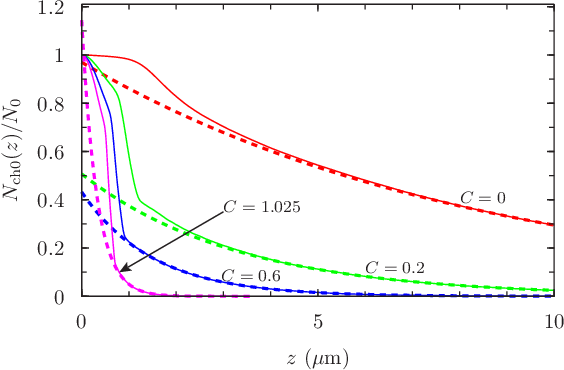}
}
\caption{The fraction $N_\mathrm{ch0}(z)/N_{0}$ of the particles that did not 
cross the channel boundaries between their entrance into the crystal and the 
point $z$ for different values of the centrifugal parameter $C$. 
The dashed lines show the corresponding exponential asymptotes $\propto \exp(-z/L_\mathrm{d})$.}
\label{ch0}
\end{figure}

The fraction $N_\mathrm{ch0}(z)/N_{0}$ 
of the particles that reached the penetration depth
$z$ without crossing the channel boundary
is shown in 
figure \ref{ch0}  as function of  $z$ for a few values of $C$.
This fraction decreases rather fast and, as expected, it has an exponential  asymptotic behavior.
Surprisingly, the exponential tail is present also at $C=1.025$. Qualitatively, the behavior of the
ratio $N_\mathrm{ch0}(z)/N_{0}$ for $C=1.025$ is the same as for $C < 1$.

\begin{figure}[htp]
\resizebox{0.95\columnwidth}{!}{%
  \includegraphics{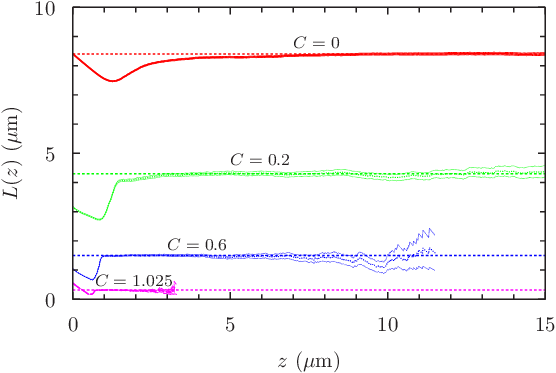}
}
\caption{The quantity $L(z)$ defined by (\ref{Lz}) that becomes equal to the dechanneling length
at large penetration depth $z$. The thin lines show the statistical errors. The straight
dashed lines correspond to the asymptotic values.}
\label{Ld}
\end{figure}

The quantity $L(z)$ defined by (\ref{Lz}) for the same values of the centrifugal parameter 
as in figure \ref{ch0}
is plotted in figure \ref{Ld}.
Indeed, $L(z)$ becomes constant (within the statistical errors) at large $z$ corresponding 
the region of the exponential behavior 
of the curves in figure \ref{ch0}. 
This is valid for $C=1.025$ as well as for $C < 1$. Only for $C>1.2$
the asymptotic region with constant values of $L(z)$ could not be identified and the 
value of $L_\mathrm{d}$ could not be found.

\begin{figure}[htp]
\resizebox{0.95\columnwidth}{!}{%
  \includegraphics{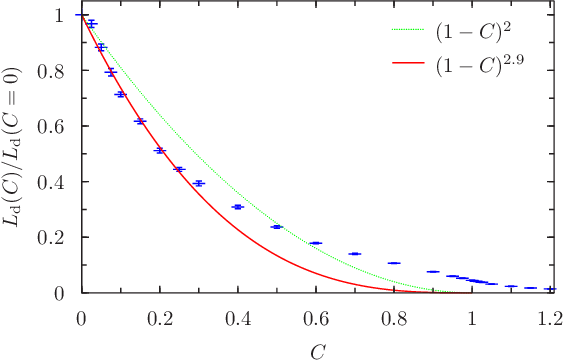}
}
\caption{The ratio $L_\mathrm{d}(C)/L_\mathrm{d}(C=0)$ of the dechanneling length 
in bent channels to the one in the straight channels as function of the centrifugal 
parameter $C$.  At $C \lesssim 0.3$ the ratio can be parametrized 
by the function $(1-C)^{2.9}$ (shown by the solid line).
For comparison, the behavior $(1-C)^2$ obtained in the diffusion
theory in the parabolic potential approximation \cite{BiryukovChesnokovKotovBook}
is shown by the dashed line.}
\label{Ld_vs_C}
\end{figure}

The dependence of the ratio $L_\mathrm{d}(C)/L_\mathrm{d}(C=0)$ on the centrifugal parameter 
is shown in figure \ref{Ld_vs_C}.

It is seen from the figure that the ratio $L_\mathrm{d}(C)/L_\mathrm{d}(C=0)$ 
follows the law $(1-C)^{2.9}$ for $C \lesssim 0.3$. At $C \gtrsim 0.3$, it decreases less steeply
and remains nonzero even at $C>1$. 

\begin{figure}[htp]
\resizebox{0.95\columnwidth}{!}{%
  \includegraphics{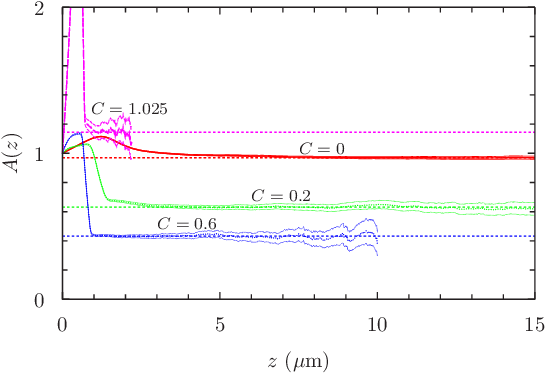}
}
\caption{
The quantity $A(z)$ defined by (\ref{Az}) that becomes equal to the asymptotic acceptance of the 
channel at large penetration depth $z$. 
The thin lines show the statistical errors. The straight
dashed lines correspond to the asymptotic values $A_\mathrm{d}$.
}
\label{Ad}
\end{figure}

The quantity $A(z)$ defined by (\ref{Az}) is plotted in figure \ref{Ad}. This quantity indeed
approaches a constant value  in the same region of $z$ where the 
quantity $L(z)$ does.  By definition, this value is the asymptotic acceptance $A_\mathrm{d}$ of the channel for 
an ideally parallel beam. Qualitatively, the behavior of the curve in the asymptotic region is the same 
for $C=1.025$ as for $C < 1$.
%(up to the statistical errors)

\begin{figure}[htp]
\resizebox{0.95\columnwidth}{!}{%
  \includegraphics{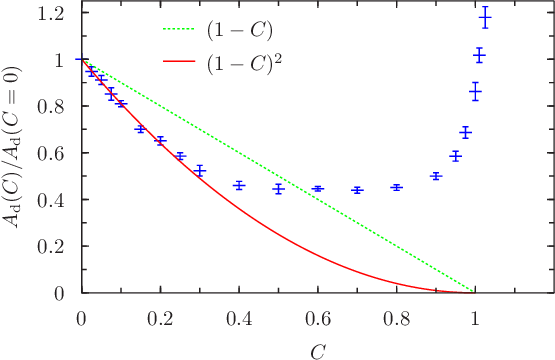}
}
\caption{
The ratio $A_\mathrm{d}(C)/A_\mathrm{d}(C=0)$ of the asymptotic acceptance of  
bent channels to the one in the straight channel as function of the centrifugal 
parameter $C$.  At $0 \lesssim C \lesssim 0.3$ the ratio can be parametrized 
by the function $(1-C)^{2}$ (shown by the solid line).
For comparison, the behavior $(1-C)$ obtained in the diffusion
theory in the parabolic potential approximation \cite{BiryukovChesnokovKotovBook} 
is shown by the dashed line.}
\label{Ad_vs_C}
\end{figure}

The dependence of the ratio $A_\mathrm{d}(C)/A_\mathrm{d}(C=0)$ on the centrifugal parameter 
is shown in figure \ref{Ad_vs_C}.
It is seen from the figure that the behavior of the ratio is changed at $C=0.3$:
it falls as $(1-C)^2$ at $0 \lesssim C \lesssim 0.3$ then it becomes nearly constant at 
$0.3 \lesssim C \lesssim 0.8 $. Finaly, the ratio starts to grow at 
$C \gtrsim 0.8 $
and becomes larger than unity at 
$C>1$. The last four points lie outside the the upper bound of the vertical axis and, 
therefore, they are not shown in the figure. Because the value of $L_\mathrm{d}$ is necessary
for calculation of $A(z)$ (see (\ref{Az})), the asymptotic acceptance $A_\mathrm{d}$ could 
not be found for $C>1.2$.

\begin{figure}[htp]
\resizebox{0.95\columnwidth}{!}{%
  \includegraphics{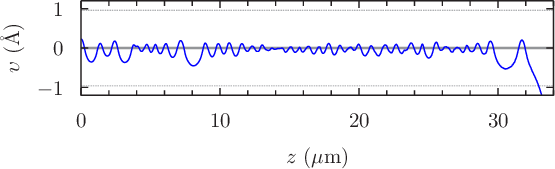}
}
\resizebox{0.95\columnwidth}{!}{%
  \includegraphics{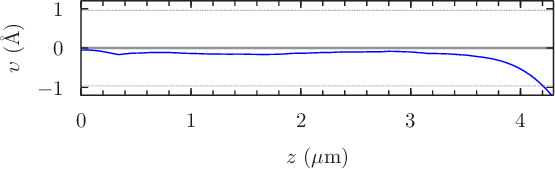}
}
\caption{
Examples of simulated trajectories for $C=0.2$ (upper panel) and $C=1.025$
(lower panel). In each case, the trajectory having the largest channeling 
segment was chosen among thousands of simulated ones. 
The vertical coordinate is $\upsilon=y-y_\mathrm{B}(z)$, cf. (\ref{upsilon}).
The thick gray lines stand for the crystal planes. The thin dashed lines 
are the channel boundaries.
}
\label{traj}
\end{figure}

Examples of trajectories with large channeling segments are shown in figure \ref{traj}.
In the case of a small centrifugal parameter, $C=0.2$, the trajectory oscillates 
around the minimum of the continuous potential. The amplitude and the phase of the 
oscillations are fluctuating due to the incoherent scattering. Nonetheless, the 
oscillation pattern is clearly seen. 

No oscillations are observed 
at a large centrifugal parameter, $C=1.025$. Instead, the particle is being 
randomly scattered. The channeling segment of a trajectory can be 
much larger than average, if random scatterings happen to keep the particle 
in the region of the gentlest slope of the centrifugally modified
continuous potential $U_\mathrm{c.f.}(\upsilon)$ defined by
(\ref{Ucf}) (cf. figure \ref{PlotUcf}).
Exactly this case is shown in the lower panel of figure \ref{traj}. 

The cases $C<1$ and $C>1$ seem to look completely different if one thinks about them
in terms of the continuous potential approximation. Indeed, the channeling 
of the particle would continue indefinitely long at $C<1$ if one considers the continuous
potential and neglects the incoherent collisions. In contrast, all the particles 
would dechannel quickly if one uses the same approximation at $C>1$. The effect of the 
incoherent collisions seems to be opposite in these two cases. They result 
into dechanneling of the projectiles at $C<1$.  On the other hand, the incoherent 
scattering prevents a substantial fraction of projectiles from immediate dechanneling
at $C>1$.
These projectiles form the exponential tail that is seen in figure \ref{ch0} for 
$C=1.025$.

Surprisingly, there is no abrupt transition between these two seemingly different 
pictures. The behavior of the curves in figures \ref{ch0}, \ref{Ld} and \ref{Ad}
is qualitatively the same for $C<1$ and $C>1$. No singularity is seen 
at $C=1$ in figures \ref{Ld_vs_C} and \ref{Ad_vs_C}. Separation of the projectile
scattering into coherent and incoherent contributions that led us to two different 
pictures seems to be misleading in this situation.

One might wonder whether the term 'channeling' is eligible in the case $C>1$. 
Indeed, there is no potential barriers between the channels. Therefore, 
the notion of channel becomes purely geometrical rather than physical.
Still, there is a measurable physical effect that legitimize using the word
channeling in the 'supercritical' regime $C>1$. 

The transverse momentum of projectile $p_{y}$ averaged over all
simulated trajectories, regardless whether they are channeling or dechanneled, is plotted
in figure $p_{y}$ versus the longitudinal coordinate $z$. It is seen that a bent crystal
preserves its guiding properties even at $C>1$. Moreover, a small steering effect is 
present even at $C=1.5$, although the values of $L_\mathrm{d}$ and $A_\mathrm{d}$ could not 
be determined in this case.

\begin{figure}[htp]
\resizebox{0.95\columnwidth}{!}{%
  \includegraphics{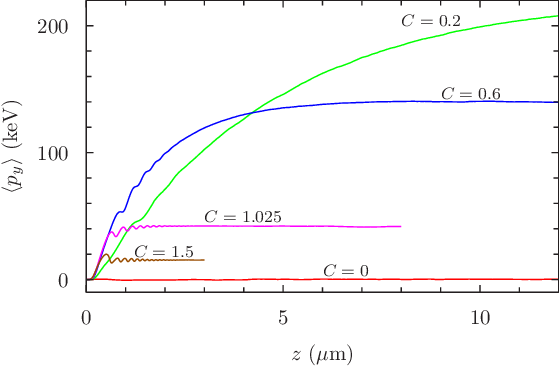}
}
\caption{The transverse momentum $p_{y}$ of the projectile averaged over all simulated
trajectories as function of the penetration depth $z$ for different values of
the centrifugal parameter $C$. 
}
\label{py}
\end{figure}

At the same time, one has to conclude that the 'supercritical channeling' is not the
most interesting case from the practical point of view.  
Electron beams can be steered more effectively using bent crystal channels with moderate 
values of the centrifugal parameter $C=0.2$--$0.3$.

\section{Conclusion and Discussion}

A Monte-Carlo study of electron channeling in the $(110)$ channel of a bent silicon
crystal is presented. The calculations were
done for the beam energy of 855 MeV for a number of values of the centrifugal parameter.

According to the simulation results, the fraction of channeling 
electrons decreases rapidly with the penetration depth $z$ and quickly approaches 
an exponential asymptote.
Similar behavior was previously seen for straight channels \cite{straight}.
This result is consistent with the one obtained 
in the kinetic theory of 
channeling in the case of positively charged projectiles
\cite{BiryukovChesnokovKotovBook}. 

A definition of the asymptotic acceptance of the channel $A_\mathrm{d}$ that is 
suitable for application within the Monte Carlo approach is formulated. The 
dependence of $A_\mathrm{d}$ on the centrifugal parameter is studied.

The dechanneling length for a set of values of the centrifugal parameter is calculated.
It decreases as $\sim (1-C)^{2.9}$ at $0 < C \lesssim 0.3$. At $C>0.3$ it decreases
less steeply and remains nonzero even at $C > 1$.

\sloppy
The behavior of the asymptotic acceptance $A_\mathrm{d}(C)$ changes 
from decreasing, $A_\mathrm{d}(C) \propto (1-C)^2$,  at $C \lesssim 0.3$
to nearly flat behavior at $0.3 \lesssim C \lesssim 0.8$. Then it starts to increase at 
$C \gtrsim 0.8$. 

It is shown that a bent crystal preserves its steering properties even 
at 'supercritical' values of the centrifugal parameters $C > 1$. However, the guiding
effect is stronger, if a bent crystal with a moderate value of $C$ is used.

It can be concluded from the above properties of electron channeling in the bent crystal that 
the period of the crystalline undulator should not exceed
several microns if one would like to observe  the undulator effect using the 
electron beam with the energy in the range of hundreds of megaelectronvolt. 
The centrifugal parameter should not be larger than $C=0.2$--$0.3$, otherwise
almost all projectiles will dechannel before the completion of the first
undulator period and, therefore, will not contribute to the undulator radiation.\footnote{The
results of the present study are relevant if the undulator period is longer than
the period of the channeling oscillations. For a very short bending period 
(in the range of hundreds of nanometers), the channeling is possible even for
very large values of the centrifugal factor $C$, provided that the bending 
amplitude is smaller than the channel width \cite{Kostyuk:2012mk}.}

The obtained result may be also useful for choosing the optimum bending of the crystals 
used for deflection and collimation of electron beams.

\section*{Acknowledgement}
At its initial stage, the work was done in the Frankfurt Institute
for Advanced Studies (FIAS) and was 
supported by the Deutsche Forschungsgemeinschaft (DFG).
The numerical simulations were done at the 
Center for Scientific Computing of 
the J.W.~Goethe University, Frankfurt.

\end{document}